\documentclass[aps, prl, twocolumn, showpacs,amsmath,amssymb]{revtex4}
\usepackage{graphicx}
\usepackage{bm}
\begin{document}

\title{Thermodynamics of the quantum Ising model in the two-dimensional kagome lattice}
\author{Chyh-Hong Chern}
\email{chern@issp.u-tokyo.ac.jp}
\author{Mitsuaki Tsukamoto}
\affiliation{Institute for Solid State Physics, University of
Tokyo, 5-1-5 Kashiwanoha, Kashiwa, Chiba 277-8581, Japan}

\begin{abstract}
In the classical Ising model on the kagome lattice, there are
macroscopically degenerate classical states with
exponentially-decayed spin correlation.  As the singular quantum
perturbation is introduced by applying the transverse magnetic
field, the quantum ground state remains disordered, dubbed by
"disorder-by-disorder".  Here we compute the temperature
dependence of the spin susceptibility and the specific heat for
difference strength of the field to understand the effect of the
singular quantum perturbation in this system.  The relevance to
the ZnCu$_3$(OH)$_6$Cl$_2$ will be also commented.
\end{abstract}

 \maketitle

Macroscopically degenerate states often occur in the classical
level in the anti-ferromagnetic systems that the exchange energy
can not be perfectly shared by all the nearest neighbor spins. It
happens naturally if spins are placed in the triangular geometry
in the two-dimensional lattices.  In two dimensions, triangles can
share the edges to form the triangular network or share the
corners to form the kagome lattice.  If we consider the classical
Ising dynamics in these two systems, both of them show
macroscopically degenerate classical ground states.  However, the
spin correlation in these two systems are very different, which
determines the fate of the consequent quantum ground states when
the quantum dynamics is introduced.  When the transverse magnetic
field is introduced by the following Hamiltonian,
\begin{eqnarray}
H=J\sum_{<ij>}S^z_i S^z_j - \Gamma \sum_i S^x_i,
\label{eq:hamiltonian1}
\end{eqnarray}
where $S^k=\sigma^k/2$ and $\sigma^k$ are the Pauli spin matrices
and $J$ is taken to be positive, the quantum ground state in the
triangular lattice favors an spin-ordered ground state, the
maximally-flippable state, because the spin correlation is
critical in the classical model.  On the other hand, the quantum
ground state in the kagome case remains disordered because the
spin correlation is exponentially-decayed\cite{moessner2000prl,
moessner2001prb}.

In this paper, we focus on the kagome case and compute its
thermodynamic properties with various $\Gamma$.  At $\Gamma=0$,
the spin susceptibility shows dramatic upturn in the low
temperature and \emph{diverges} at the zero temperature.  For
non-zero $\Gamma$, we also see the significant upturn but it
\emph{saturates} at the zero temperature.  Our calculation is the
first results on the thermodynamic quantities to distinguish the
classical disordered state and the quantum disorder states.
Furthermore, as the macroscopically degenerate classical states
lead to the residual entropy at $T=0$, the quantum dynamics which
lifts the degeneracy results in additional peak in the specific
heat in the low temperature at the order of $\Gamma$.  Due to
these properties, our model might be relevant to the recent
discovered ZnCu$_3$(OH)$_6$Cl$_2$, which shows the abnormal upturn
in the spin susceptibility and saturates at $T=0$.  There have
been several theoretical papers trying to describe this unusual
property.  A straightforward interpretation is the presence of the
impurity\cite{ran2007prl} that contributes additionally to the
susceptibility.  However, a naive inclusion of the contribution
from the impurity can explain the data only above
20K\cite{misguich2007}. Different angles of views by considering
the contribution from the spin anisotropy and spin-orbital
interaction can somewhat show the upturn but can not show the
saturation at $T=0$ due to the "minus-sign" problem in their
calculation\cite{rigol2007prb, rigol2007prl}.  Our calculation
advances their results because there is no "minus-sign" problem in
our model.  Moreover, our phenomenological approach to understand
the magnetic systems also opens a new way to explore the
low-temperature properties that can never be reached by the
present quantum Monte Carlo technique due to the technical
difficulty.

In the following, we use the quantum Monte Carlo technique to
compute the thermodynamical quantities of
Eq.(\ref{eq:hamiltonian1}) by using the Trotter-Suzuki
approximation \cite{suzuki1976}.  To calculate the magnetic
susceptibility, we have to include the perturbed Hamiltonian
$\delta H = -h\sum_i S^z_i - h_y\sum_i S^y_i$. Discreting the
imaginary time direction by $n$ steps, the n$^{\text{th}}$
Trotter-Suziki approximant of the partition function is given by
\begin{eqnarray}
&&Z^{(n)}=\sum_{\{\sigma_{jk}\}}e^{-H'_{\text{eff}}}[\cosh\frac{\beta
R}{2n}]^{Nn},
\nonumber\\
&&H'_{\text{eff}}=\frac{\beta
J}{4n}\sum_{k,<ij>}\sigma_{ik}\sigma_{jk}-\frac{\beta
h}{2n}\sum_{j,k}\sigma_{jk}\nonumber\\&&-\frac{1}{2}\log\coth\frac{\beta
R }{2n}\sum_{j,k}(\sigma_{jk}\sigma_{j,k+1}-1)
\label{eq:trotter-suzuki}
\end{eqnarray}
where $R^2=\Gamma^2+h^2_y$, $N$ is the number of the lattice
sites, $\sigma_{jk}$ are now the classical variables taking only
+1 and -1, and the summation is over the $n$ stacks of the kagome
lattices. The cluster algorithm is applied along the imaginary
time direction. The thermodynamical quantities can be obtained by
taking the derivatives on Eq.(\ref{eq:trotter-suzuki}) with
respect to the corresponding thermodynamical variables.  The
results are summarized as the following:
\begin{eqnarray}
&&Cv^{(n)}T^2=\!\frac{1}{N}\!<\!(\frac{\partial
H'_{\text{eff}}}{\partial
\beta})^2\!-\!\frac{\partial^2H'_{\text{eff}}}{\partial
 \beta^2}\!>\!-\!\frac{1}{N}<\!\frac{\partial
H'_{\text{eff}}}{\partial \beta}\!>^2\nonumber\\&&\ \ \ \ \ \ \ \
\ \ \ \ \ \ +\frac{R^2}{n\cosh^2\frac{\beta R}{n}}
\label{eq:cv}\\&&\chi^{(n)}_{zz}=\frac{T}{N}(<(\frac{\partial
H'_{\text{eff}}}{\partial
 h})^2>-<\frac{\partial
H'_{\text{eff}}}{\partial h}>^2) \label{eq:chzz}\\
&&\chi^{(n)}_{yy}= \frac{T}{N}(<(\frac{\partial
H'_{\text{eff}}}{\partial
h_y})^2-\frac{\partial^2H'_{\text{eff}}}{\partial
 h_y^2}>-<\frac{\partial H'_{\text{eff}}}{\partial
h_y}>^2)\nonumber \\
&&\ \ \ \ \ \ \ \ \ \ +\frac{h_y^2}{nTR^2\cosh^2\frac{\beta
R}{n}}+\frac{\Gamma^2}{R^3}\tanh\frac{\beta R}{n}
\label{eq:chyy}\\&&\chi^{(n)}_{xx}= \frac{T}{N}(<(\frac{\partial
H'_{\text{eff}}}{\partial
\Gamma})^2-\frac{\partial^2H'_{\text{eff}}}{\partial
 \Gamma^2}>-<\frac{\partial H'_{\text{eff}}}{\partial
\Gamma}>^2)\nonumber \\
&&\ \ \ \ \ \ \ \ \ \ +\frac{1}{nT\cosh^2\frac{\beta R}{n}}
\label{eq:chxx}
\end{eqnarray}
where $\chi_{zz}$, $\chi_{yy}$, and $\chi_{xx}$ are defined by $d
m_z/ dh$, $d m_y / dh_y$ and $d m_x/ d\Gamma$ respectively, and
$m_{x,y,z}$ are the magnetization per site.  For a fixed
temperature, the leading term of the error between the
Trotter-Suzuki approximant of the thermodynamical quantities and
those obtained from Eq.(\ref{eq:hamiltonian1}) can be shown to be
linear in $1/n$. Therefore, our results are the extrapolation of
the straight line at $n\rightarrow \infty$. For each $n$ and an
ensemble, $10^6$ Monte Carlo Sweeps are taken. For each set of
parameters, 64 ensemble averages are used.

\begin{figure}[htb]
\includegraphics[width=0.5\textwidth]{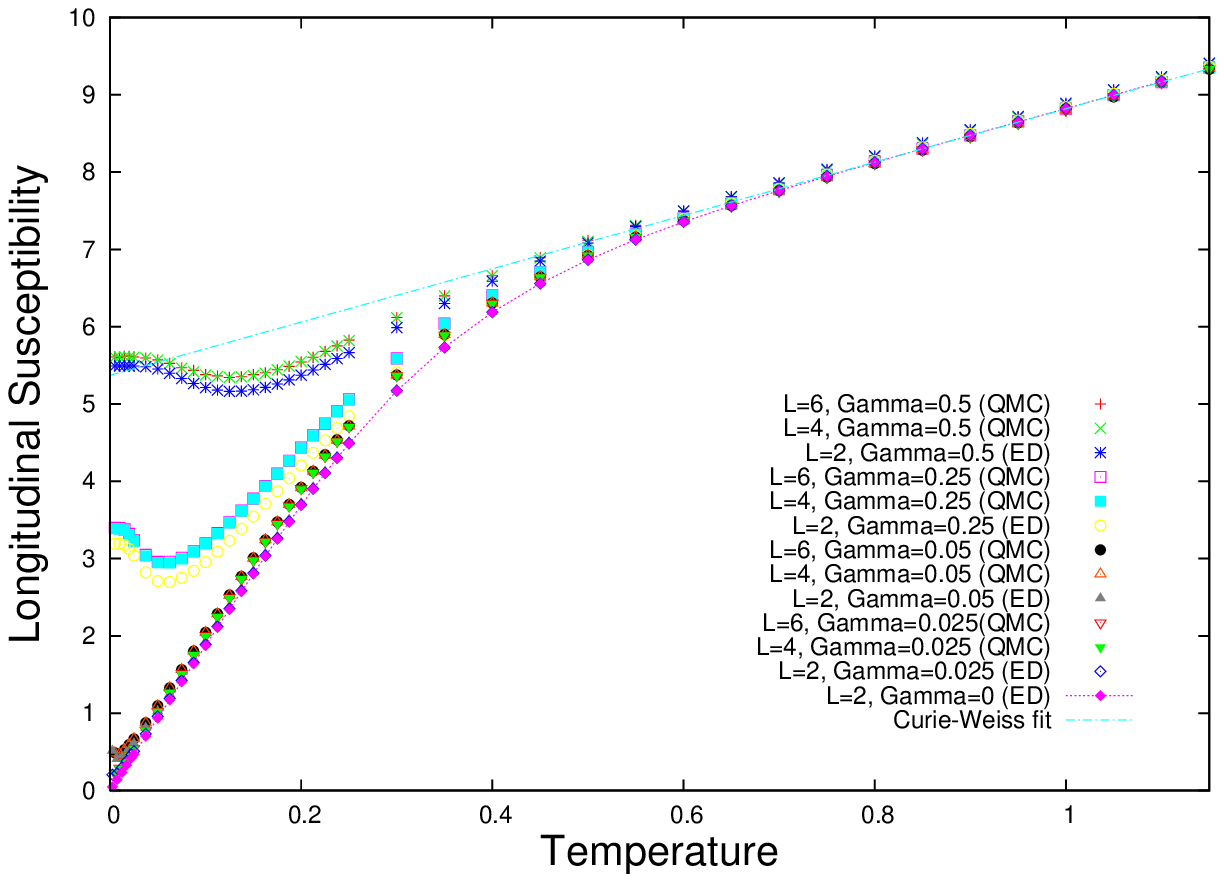}
\caption{(Color Online)The temperature (in the unit of $J$)
dependence of the inverse of the $\chi_{zz}$ per spin. L=2, 4, 6
correspond to N = 12, 48, 108 respectively. The blue straight line
is the Curie-Weiss fit. QMC is denoted for the result from quantum
Monte Carlo and ED is the one from exact
diagonalization}\label{Fig:chzz}
\includegraphics[width=0.5\textwidth]{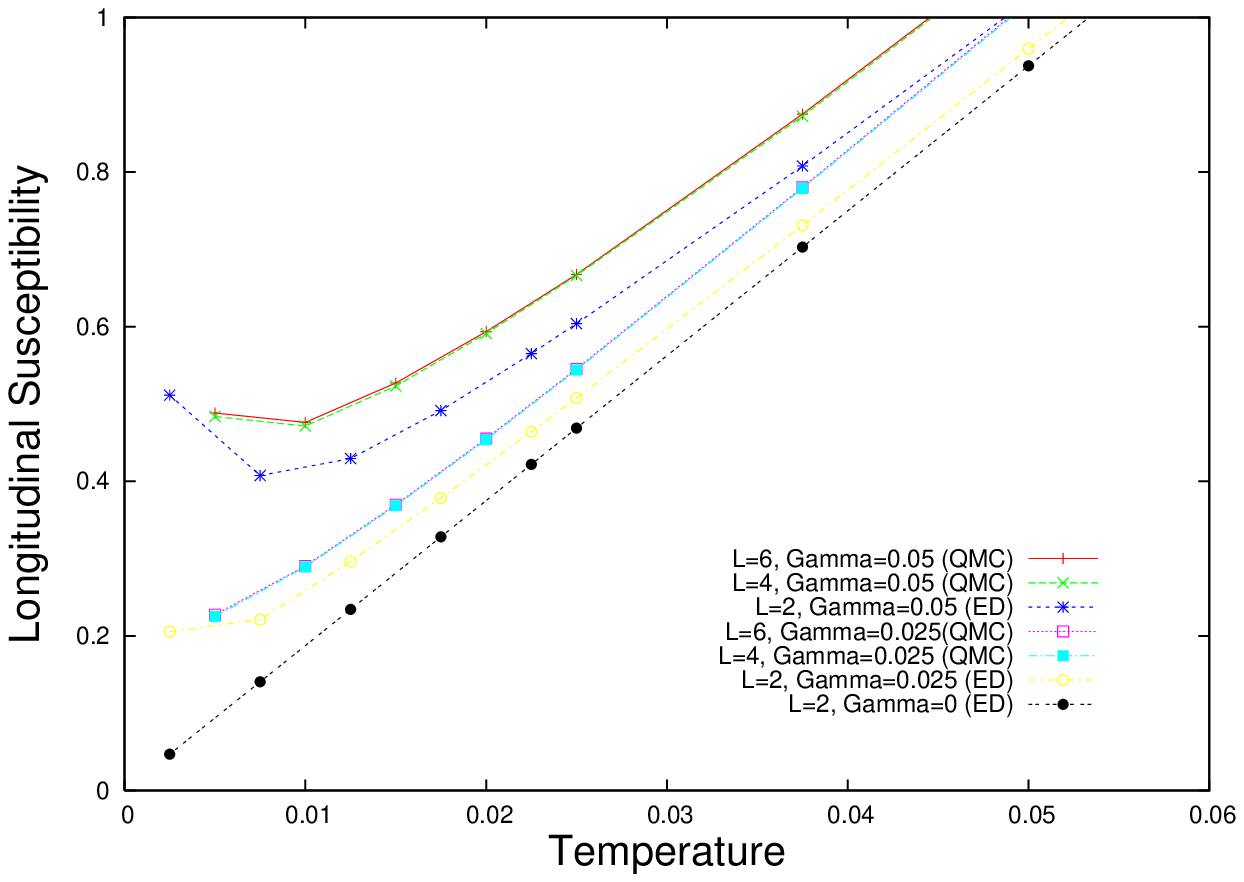}
\caption{(Color Online)The magnifying plot around the origin of
Fig.(\ref{Fig:chzz}). At $\Gamma=0$ (classical disorder), the
longitudinal susceptibility diverges at $T=0$. At finite $\Gamma$
(quantum disorder) , they saturate at finite values at $T=0$.
(Temperature is in the unit of $J$)} \label{Fig:zchzz}
\end{figure}

Fig.(\ref{Fig:chzz}) and (\ref{Fig:zchzz}) are the temperature
dependence of the inverse of the longitudinal susceptibility
$\chi_{zz}$ per spin for various $\Gamma$ and different sizes of
the system. $L$ in the figures is the length of the linear
dimension so that the number of total spins $N=3L^2$. The period
boundary condition is applied in both directions.  The temperature
is in the unit of J and the Boltzman constant is set to be 1.  In
high temperature, all curves are fitted well by the Curie-Weiss
law regardless $\Gamma$ and $L$. As the temperature goes down, the
$\chi_{zz}$ have the abnormal upturn and saturate at the values
depending on the $\Gamma$ at $T=0$. Spin flipping helps reduce the
upturn of the $\chi_{zz}$. Since our Monte Carlo results for $L=2$
is the same as the ones obtained by the exact diagonalization, we
only report the exact diagonalization results for $L=2$ here.  For
$\Gamma=0$, the coupling constant in the imaginary-time direction
is divergent so that we report the result of $L=2$ by the exact
diagonalization only.  At finite $\Gamma$, the ground state is
quantum disorder. $L=4$ looks fairly enough for the calculation
and the size dependence becomes weaker for smaller $\Gamma$.
Moreover, since the ground state of the $\Gamma=0$ is known as a
classical disorder state \cite{moessner2000prl}, we believe that
the result of $L=2$ for $\Gamma=0$ is representative.
Fig.(\ref{Fig:zchzz}) is the magnifying plot around the origin of
the Fig.(\ref{Fig:chzz}). It can be seen clearly that the
$\chi_{zz}$ of the $\Gamma=0$ (classical disorder) diverges at
$T=0$, while the ones at the finite $\Gamma$ (quantum disorder)
saturate at $T=0$.  This is the first result of the magnetic
susceptibility thus far that distinguishes the quantum disorder
from the classical disorder.

\begin{figure}[htb]
\includegraphics[width=0.5\textwidth]{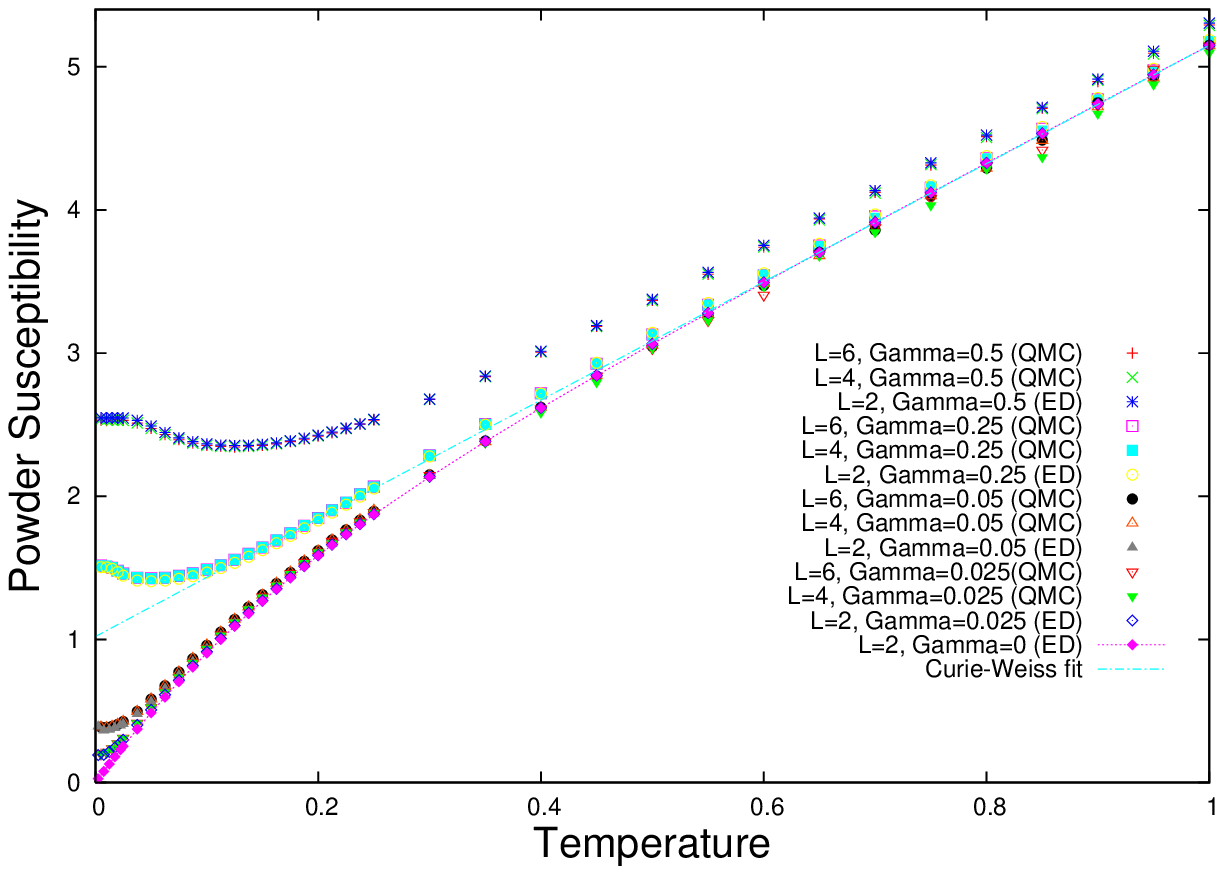}
\caption{(Color Online)The temperature dependence of the inverse
of the powder susceptibility $\chi$ per spin. L=2, 4, 6 correspond
to N = 12, 48, 108 respectively. The blue straight line is the
Curie-Weiss fit. QMC is denoted for the result from quantum Monte
Carlo, and ED is the one from exact diagonalization. (Temperature
is in the unit of $J$.)}\label{Fig:sus}
\includegraphics[width=0.5\textwidth]{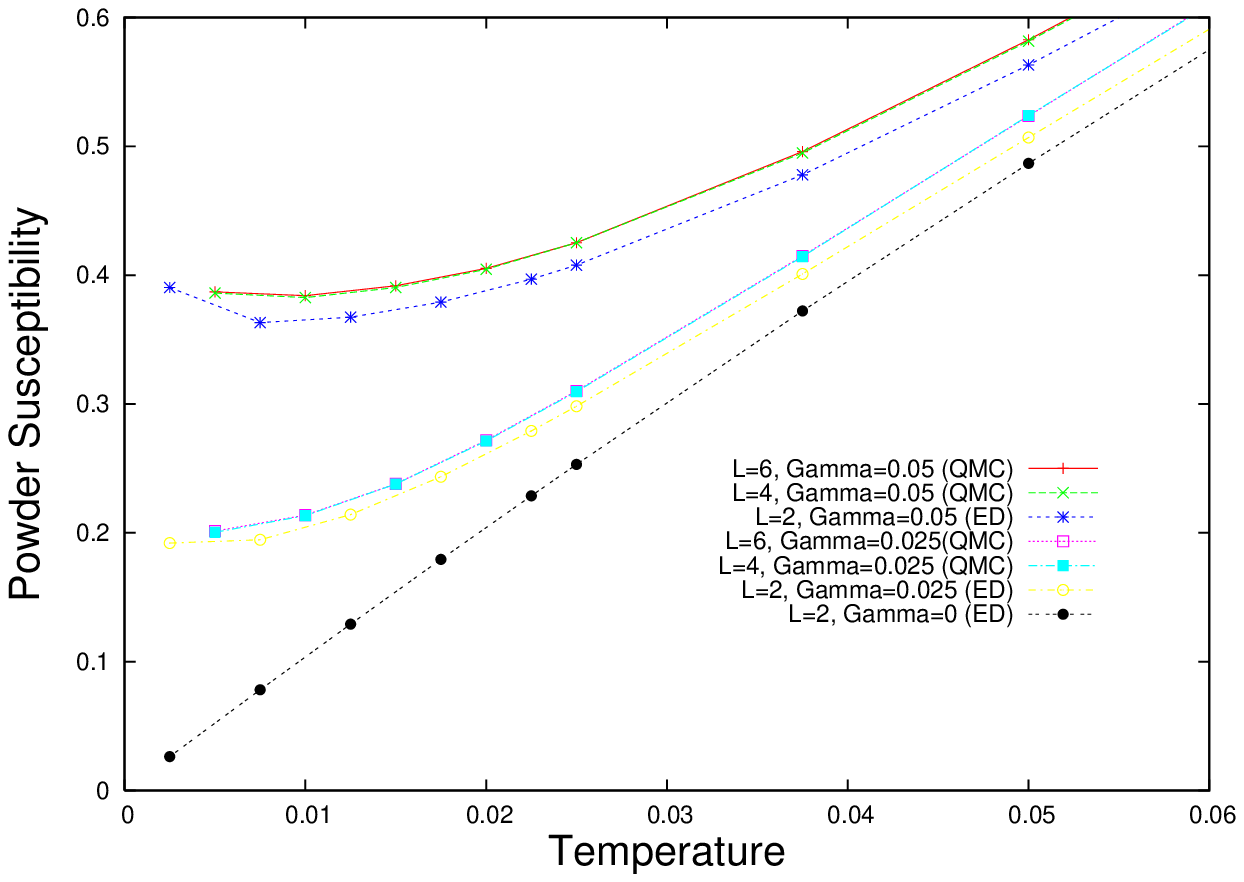}
\caption{(Color Online)The magnifying plot around the origin of
Fig.(\ref{Fig:sus}).  At $\Gamma=0$ (classical disorder), the
powder susceptibility diverges at $T=0$. At finite $\Gamma$
(quantum disorder), they saturate at finite values at $T=0$.
(Temperature is in the unit of $J$)} \label{Fig:zsus}
\end{figure}

Experiments measured the magnetic susceptibility in the powder
samples.  In Fig.(\ref{Fig:sus}) and (\ref{Fig:zsus}), we report
the temperature dependence of the inverse of the powder
susceptibility $\chi$ defined by
$\chi=1/3(\chi_{xx}+\chi_{yy}+\chi_{zz})$. One apparent feature is
that the size scaling becomes even weaker. It is probably because
$\chi$ is an averaged quantity.  Because of the anisotropy,
$\chi_{xx}$, $\chi_{yy}$, and $\chi_{zz}$ are different from one
other. Surprisingly, $\chi_{yy}$ is larger than the others and the
size scaling is small in $\chi_{yy}$.  On the other hand,
$\chi_{yy}$ and $\chi_{xx}$ become equal at $\Gamma=0$, which is
consistent with expectation. Furthermore, the $\chi$ shows the
upturn in the low temperature and saturates at the zero
temperature for the small $\Gamma$ cases.  In
Fig.(\ref{Fig:zsus}), the difference between classical disorder
and the quantum disorder is again clearly seen. The $\chi$ of the
quantum disorder saturates at $T=0$, and the one of the classical
disorder diverges.

\begin{figure}[htb]
\includegraphics[width=0.5\textwidth]{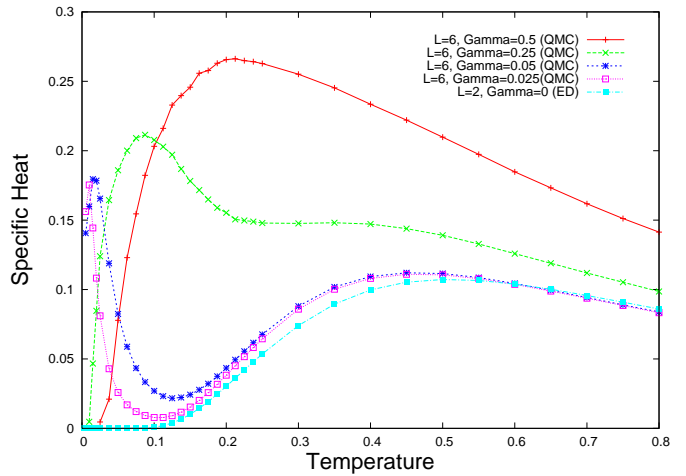}
\caption{(Color online)The temperature (in the unit of $J$)
dependence of specific heat per spin. L=6 corresponds to N=108.
QMC is denoted for the result from quantum Monte Carlo, and ED is
the one from exact diagonalization}\label{Fig:cv}
\end{figure}

In Fig.(\ref{Fig:cv}), we report the temperature dependence of the
specific heat with various $\Gamma$.  At $\Gamma=0$, a broad peak
locating at the order of $J$ is clearly observed.  In this case,
there is also a $\delta-$function peak at $T=0$ representing the
macroscopic degeneracy of the classical ground states.  At finite
$\Gamma$, the transverse field lifts the degeneracy and therefore
results in an additional peak locating at the order of $\Gamma$.
For larger $\Gamma$, like 0.25 and 0.5, two peaks emerges and a
spin gap can be clearly seen. For $\Gamma=0.025$ and 0.05, two
peaks are well separated.  The one at the order of $J$ at these
two $\Gamma$s are overlapping with the one at $\Gamma=0$. This
indicates that the quantum perturbation has the dramatic effect on
the ground states but not the higher-energy states. Unfortunately,
our numerical resolution is not able to identify whether or not it
is gapped or gapless for small but non-vanishing $\Gamma$. We
remark that the states in the degenerate manifold at $\Gamma=0$
are connected by the single spin flip, \emph{i.e.} gapless.
Whether or not the gaplessness survives with respect to the
quantum perturbation and extends to finite $\Gamma$ requires more
detail study.

For the application to ZnCu$_3$(OH)$_6$Cl$_2$, there is only one
parameter in our model to fit with the experiments.  Although the
$\Gamma$ term is understood as the external transverse magnetic
field, it can also be regarded as the internal quantum spin
flipping matrix.  In addition, at a first glance, considering the
Ising dynamics in the spin-1/2 system might be absurd for most of
the experts. However, it has been shown that even if the spin is
isotropic in the bulk of Fe, Ni, and V, the effective spin
interaction becomes anisotropic when the dimension is reduced,
namely thin film, as long as there is non-vanishing spin-orbital
interaction \cite{gay1986prl}. Therefore, that the spin in the
spin-1/2 system is isotropic in 3 dimensions does not guarantee in
2 dimensions. In the presence framework, the saturation seen in
the experiments\cite{imai2007, ofer2007} can be easily understood.
Namely, the quantum spin flip process dominates the low-energy
physics as long as the Ising dynamics is relevant.

Our proposal can be tested experimentally by measuring the
magnetic susceptibility on the single crystal sample of
ZnCu$_3$(OH)$_6$Cl$_2$.  If the spin flip process is relevant,
experiments should see the shift of the saturation as the
transverse magnetic field varies.  Because the spin flip dynamics
is enhanced, we expect the saturation shifts down as the field
increases.  In the very recent manuscript \cite{bert2007}, this
effect is observed in the powder sample.  The applied magnetic
field has two components: one is parallel to the hard plane and
the other is longitudinal to the easy axis.  Because the
susceptibility is not sensitive to the longitudinal field, the
decreasing saturation value with increasing applied field can be
understood as the enhancement of the spin flipping process by the
transverse field.

In summary, we compute the temperature dependence of the spin
susceptibility and the specific heat of the quantum Ising model.
Different from the Heisenberg model in the kagome lattice, the
spin susceptibility in our model has the intrinsic upturn and
saturation at finite value.  In the specific heat, we observe the
two-peak structure and the existence of the spin gap for larger
$\Gamma$.  The finite temperature properties in
ZnCu$_3$(OH)$_6$Cl$_2$ indicates that the spin anisotropy and
spin-flipping process might be dominant in the low temperature. In
this case, the robust spin dynamics persists to the zero
temperature might be the reason that ZnCu$_3$(OH)$_6$Cl$_2$
remains disordered at the mK range.  In addition, there is no
minus-sign problem in our model even though kagome is a highly
frustrated system, which advances other methods technically to
study the low-energy properties of the ZnCu$_3$(OH)$_6$Cl$_2$.

CHC is indebted to Naoto Nagaosa and Masaki Oshikawa for very
fruitful discussion. Part of the calculation is done by using the
facilities in the Supercomputer Center in the Institute for Solid
State Physics.


\end{document}